\documentclass[review]{elsarticle}
\usepackage{amsmath,amssymb}
\usepackage{graphicx}
\usepackage{subcaption}
\usepackage{booktabs}
\usepackage[colorlinks=true,citecolor=blue,linkcolor=blue,urlcolor=blue]{hyperref}
\usepackage{caption}

\def\kv{{\bf k}}

\def\pv{{\bf p}}

\newcommand{\bra}[1]{\left\langle{#1}\right|}
\newcommand{\ket}[1]{\left|{#1}\right\rangle}

\newcommand\vecty[3]{\ensuremath{%
		\edef\TEMPnum{#1#2#3}\def\TEMPzero{000}%
		\ifx\TEMPnum\TEMPzero 0\else%
		\ifcase#1\or\hat{i}\ifnum#2>0 +\fi
		\else#1\hat{i}\ifnum#2>0 +\else\ifnum#3>0+\fi\fi\fi
		\ifcase#2 \or\hat{j}\ifnum#3>0 +\fi
		\else #2\hat{j} \ifnum#3>0 + \fi\fi
		\ifcase#3 \or\hat{k}
		\else#3\hat{k}\fi%
		\fi}}
\journal{Journal of Physics: Condensed Matter}
\begin{document}
	\begin{frontmatter}
		\title{Symmetry-constrained low-energy effective Hamiltonian for topological RuC and OsC monolayers}
		
		\author[inst1]{A.Baradaran\corref{cor1}}
		\cortext[cor1]{Corresponding author}
		\ead{a.baradaran@ut.ac.ir}
		
		\author[inst2]{O. Sedaghatfar}
		
		\address[inst1]{Department of Physics, University of Qom, Qom, Iran}
		\address[inst2]{Department of Mathematics,YI.C., Islamic Azad University, Tehran, Iran}
		
\begin{abstract}
	We derive a low-energy $\mathbf{k}\cdot\mathbf{p}$ effective Hamiltonian for monolayer osmium carbide (OsC) and ruthenium carbide (RuC) in a planar hexagonal configuration. First-principles calculations indicate that both monolayers are dynamically stable and exhibit features of a two-dimensional quantum spin Hall (QSH) phase, characterized by a nontrivial $\mathbb{Z}_2$ topological invariant. Using symmetry analysis at the $\Gamma$ point, we construct a multiband $\mathbf{k}\cdot\mathbf{p}$ Hamiltonian including spin-orbit coupling and reduce it to a four-band low-energy model through Löwdin partitioning. The effective Hamiltonian has a block-diagonal form, with two blocks related by time-reversal symmetry, analogous to the Bernevig--Hughes--Zhang (BHZ) model. In contrast to the standard BHZ form, the symmetry-allowed off-diagonal coupling contains quadratic momentum-dependent terms, which modify the low-energy dispersion near the $\Gamma$ point. The fitted parameters reproduce the ab initio band structures in the low-energy region, yielding a compact model for analyzing the electronic and topological properties of monolayer OsC and RuC.
\end{abstract}
		
	\end{frontmatter}
	
\section{Introduction}\label{intro}

Two-dimensional topological insulators (2D TIs), also known as QSH insulators, host spin-polarized helical edge states protected by time-reversal symmetry~\cite{Roth2009}. These states support dissipationless charge and spin transport, making 2D TIs attractive for spintronic, low-power electronic, and quantum-device applications~\cite{Kou2017,Weber_2024}. However, the realization of robust QSH states at room temperature remains challenging, motivating the search for new materials with sizable topological band gaps~\cite{Zhang2021,Weber_2024,Kou2017,baradaran_2020}.

The QSH phase was first proposed for HgTe/CdTe quantum wells within the BHZ model~\cite{Bernevig2006} and later confirmed experimentally in HgTe and InAs/GaSb quantum-well structures~\cite{konig2007,kenz2011}. The discovery of graphene stimulated extensive efforts to identify atomically thin QSH materials~\cite{Kane2005,Kou2017}. However, the weak SOC in graphene and correspondingly small topological band gaps predicted for related honeycomb materials, such as silicene and germanene, have hindered the realization of robust room-temperature QSH behavior~\cite{Yao2007,Liu2011,Ezawa2015,Bampoulis2023}.

Beyond graphene-derived systems, a variety of alternative two-dimensional material platforms have been proposed and, in some cases, experimentally realized as 2D TIs, including distorted 1T$'$-WTe$_2$~\cite{Sanfeng2018,Zhao2020}, bismuthene based monolayers~\cite{Reis2017,LU2024,Kou2018}, and several transition-metal carbides belonging to the MXene family, a class of two-dimensional transition-metal carbides and nitrides~\cite{si2016}. In these materials, nontrivial topological phases arise from the interplay among crystal symmetry, orbital hybridization, and strong SOC, resulting in sizable band gaps and robust helical edge states~\cite{Weber_2024,HfHalides2024}.

Among the emerging classes of 2D TI materials, transition-metal carbides have attracted growing attention owing to their rich electronic structures and the possibility of strong SOC in compounds containing heavy transition metals~\cite{si2016,Weber_2024}. In addition to their potential for realizing nontrivial topological phases, binary transition-metal carbides exhibit excellent mechanical strength, thermal stability, and chemical robustness~\cite{zheng2005superhard,TMC:synthesis2016,TMC2D2015}.

Noble-metal carbides such as OsC and RuC are particularly interesting members of this family. These compounds have been synthesized or predicted in WC-type layered hexagonal structures and in several additional stable phases under high-pressure and high-temperature conditions~\cite{kempter1960preparation,kempter1964further,li2012predicting,fadda2017new}. Motivated by the growing interest in two-dimensional transition-metal monocarbides~\cite{Qin2021,Zhao2026}, recent first-principles studies have proposed hexagonal monolayers of OsC and RuC and demonstrated their structural and dynamical stability~\cite{bentaibi2022new,RuC-turk}. More broadly, transition-metal monocarbides have emerged as promising platforms for topological and phononic phenomena~\cite{Zhao2026,Sufyan2022}.

Although the topological properties of monolayer OsC have been explored previously~\cite{bentaibi2022new}, a symmetry-based low-energy effective Hamiltonian describing the electronic states near the Fermi level is still lacking. Moreover, a corresponding analytical description for monolayer RuC has not yet been reported. The absence of such models limits analytical investigations of the microscopic origin of the band inversion and the role of crystal symmetry in stabilizing the topological phase. While first-principles calculations provide detailed electronic structures, they do not yield a compact analytical description of the low-energy states required for studying external perturbations such as magnetic fields, strain, or electrostatic gating. Developing reliable effective Hamiltonians is therefore essential for connecting first-principles calculations with experimentally accessible low-energy phenomena.

In this work, we develop low-energy effective Hamiltonians for monolayer RuC and OsC by combining first-principles calculations, symmetry analysis, and $\mathbf{k}\cdot\mathbf{p}$ theory. Starting from spin-orbit-coupled multiband Hamiltonians, we derive effective four-band models using Löwdin partitioning. The resulting Hamiltonians consist of two time-reversal-related blocks similar to the BHZ model, while containing additional symmetry-allowed quadratic off-diagonal momentum-dependent terms originating from the $D_{3h}$ crystal symmetry, beyond the standard BHZ form. The models accurately reproduce the low-energy DFT band structures near the $\Gamma$ point and provide analytical descriptions of the spin-orbit-coupled electronic states and topological band inversion in RuC and OsC monolayers. Furthermore, they establish a framework for investigating magnetic-field-induced spectra and other low-energy topological phenomena in these materials.

The paper is organized as follows. In Sec.~\ref{Sec:Ab initio}, we present the first-principles electronic structure, including the band dispersion, symmetry representations, and orbital character of the low-energy states. In Sec.~\ref{sec:effective_hamiltonian}, we construct the multiband $\mathbf{k}\cdot\mathbf{p}$ model at the $\Gamma$ point and derive a effective $4\times4$ low-energy Hamiltonian using Löwdin partitioning, with model parameters obtained from fits to the ab initio band structure. Finally, Sec.~\ref{sec:conclusion} summarizes the main findings and discusses the physical implications and possible applications of the resulting effective model.

\section{Ab initio calculation}\label{Sec:Ab initio}

In this section, we discuss the crystal structure, electronic band structures, orbital characters, and band symmetries of monolayer OsC and RuC in the planar hexagonal structure. These results are used in the following section to construct the low-energy $\mathbf{k}\cdot\mathbf{p}$ effective Hamiltonian around the $\Gamma$ point.

\subsection{Crystal structure and symmetries}
\begin{figure}[t]
	\centering
	\begin{tabular}[c]{cc}	
		\begin{minipage}[c]{0.65\textwidth}
			\includegraphics[width=1\textwidth]{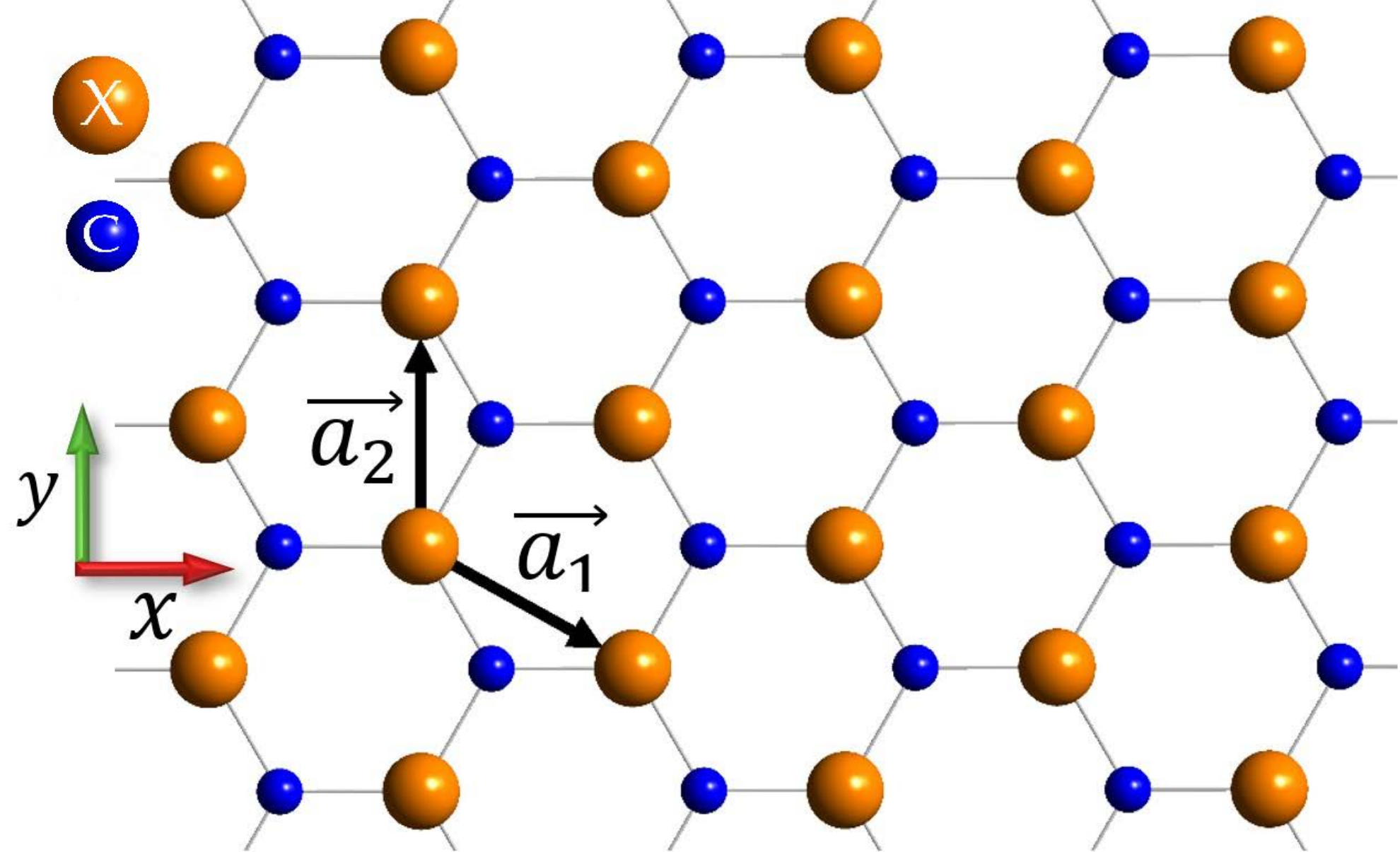}\\
			\includegraphics[width=1\textwidth]{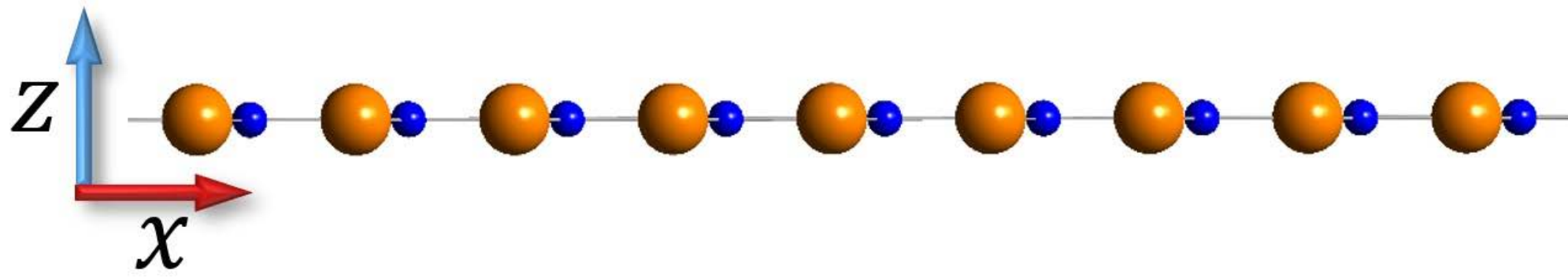}
		\end{minipage}
	\end{tabular}
	\caption{\label{fig:1}Schematic view of the monolayer XC lattice (X = Os or Ru): top view (top) and side view (bottom).}
\end{figure}

The monolayer XC structure, where X denotes Os or Ru, contains two atoms in the unit cell. The X atom is located at $(0,0,0)$, while the carbon atom is located at $(2a/3,a/3,0)$, as illustrated in Fig.~\ref{fig:1}. The primitive lattice vectors are taken as
$\vec{a}_1=\sqrt{3}a/2~\vecty{1}{0}{0}-a/2~\vecty{0}{1}{0}$ and
$\vec{a}_2=a\vecty{0}{1}{0}$.
The optimized lattice constants are approximately $3.23$~\AA{} for OsC and $3.25$~\AA{} for RuC.

Both monolayers belong to the $P\bar{6}m2$ space group and have $D_{3h}$ point-group symmetry. The corresponding symmetry operations include the identity $E$, threefold rotations $C_3$, twofold rotations $C_2'$ about in-plane axes, horizontal mirror reflection $\sigma_h$, improper rotations $S_3$, and vertical mirror reflections $\sigma_v$.

\subsection{Computational details and electronic structure}
\begin{figure*}[t]
	\centering
	\includegraphics[width=1.0\textwidth]{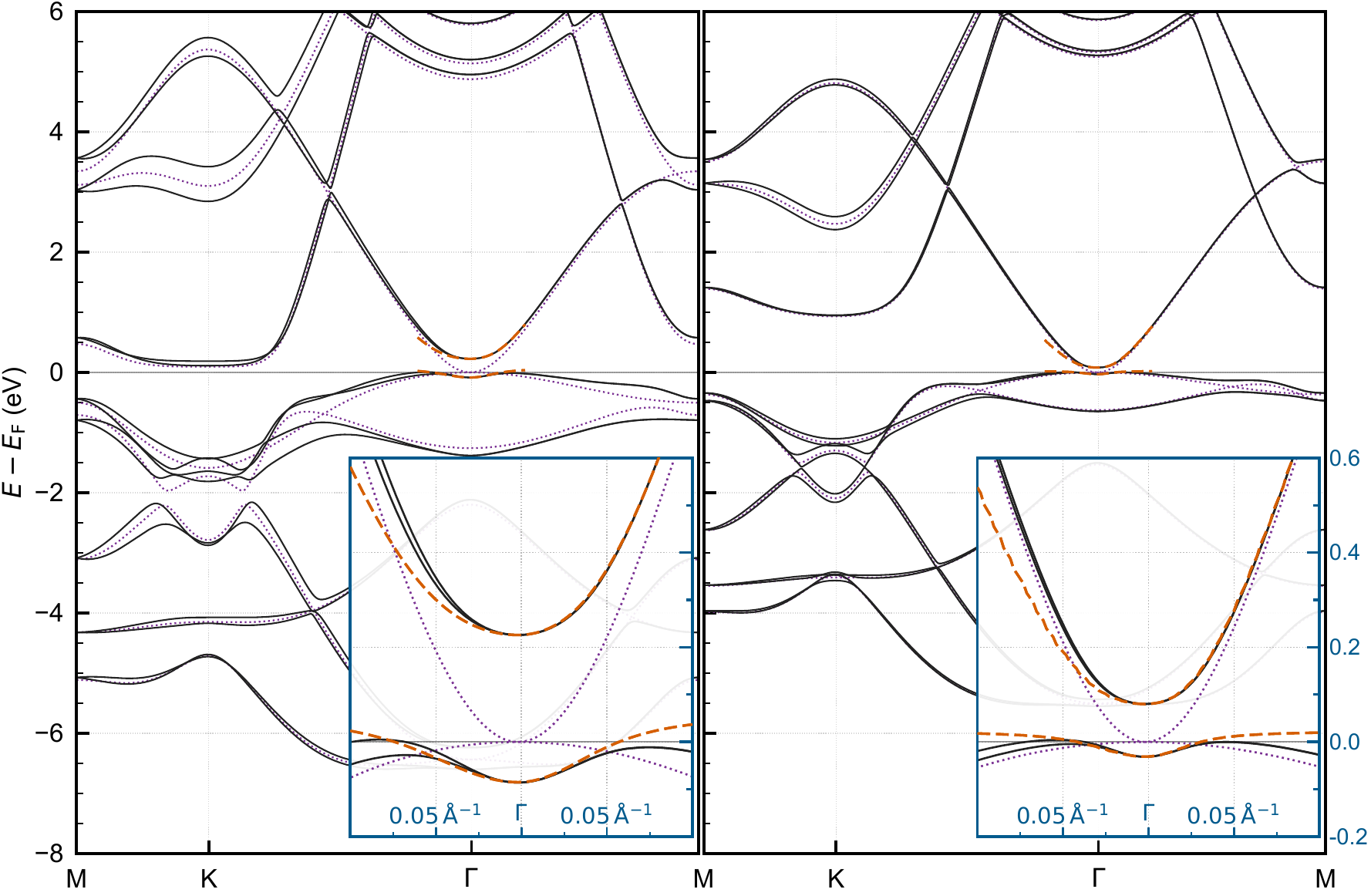}	
	\caption{\label{fig:band_structure}
		Band structures of monolayer OsC (left) and RuC (right) in the planar hexagonal structure obtained from \textit{ab initio} calculations. Solid and dotted lines denote results with and without SOC, respectively. Red dashed curves show the low-energy bands obtained from the effective Hamiltonian in Eq.~\eqref{eq:4x4ham}. Insets show the bands near the $\Gamma$ point along the $\Gamma$--K and $\Gamma$--M directions. The effective model is fitted only within the range 		$|\kv|<0.1~\mathrm{\AA^{-1}}$ for OsC and $|\kv|<0.05~\mathrm{\AA^{-1}}$ for RuC; therefore, the dashed curves are intended to reproduce the low-energy band inversion and SOC-induced gap near $\Gamma$, not the full dispersion along the high-symmetry path.}
\end{figure*}

First-principles density-functional-theory (DFT) calculations were performed within the generalized-gradient approximation (GGA) using the Wu--Cohen exchange-correlation functional \cite{WC2009,Hamann_1979_NCPS}, which improves the description of structural parameters compared with standard PBE \cite{Hammer_1999_PBE}. Equilibrium lattice parameters were obtained by fitting the total energy as a function of lattice constant to an equation of state. In the interstitial region, the charge density and potential were expanded in Fourier components up to $G_{\mathrm{max}}=12~\mathrm{Bohr}^{-1}$, corresponding to an approximate cutoff of $144$ Ry. Brillouin-zone integrations used a $31\times31\times1$ $k$-point mesh without symmetry reduction. A vacuum layer of $20$~\AA{} was included along the out-of-plane direction. The total energy and Hellmann--Feynman forces were converged to $10^{-4}$~Ry and $10^{-3}$~Ry~\AA$^{-1}$, respectively.

Fig.~\ref{fig:band_structure} shows the calculated band structures of OsC and RuC with and without SOC. Without SOC, the valence and conduction bands exhibit a quadratic band touching near the Fermi level at $\Gamma$ point. Including SOC opens a gap and produces an inversion of predominantly $d$-orbital-derived bands near
$\Gamma$. For OsC, a related band-inversion mechanism was discussed~\cite{bentaibi2022new}.

To characterize the topology of our systems, we calculated the $\mathbb{Z}_2$ invariant using the lattice Chern number method \cite{Fukui_2007_LCN}, as implemented in OpenMX \cite{OpenMX1,OpenMX2}. This method is appropriate for systems without inversion symmetry, where the parity-eigenvalue criterion is not directly applicable \cite{Fu_2007_Z2,Fu_2006_Z2}. We obtain $\mathbb{Z}_2=1$ for both OsC and RuC, confirming their two-dimensional topological-insulator character. For OsC, this result agrees with the previously reported Wilson-loop analysis \cite{bentaibi2022new}.

The energy positions of the conduction-band minimum (CBM) and valence-band maximum (VBM) at $\Gamma$, M, and K are listed in Tab.~\ref{table:energy_value}. All energies are measured relative to the Fermi level. Since GGA-based calculations generally underestimate band gaps \cite{baradaran_2022,Borlido_2020_gapVSgga}, the numerical gap values should be interpreted with this limitation in mind.

For RuC, the CBM and VBM closest to the Fermi level are both located at $\Gamma$, while the corresponding states at K and M are farther away in energy. Thus, the low-energy electronic structure of RuC is well described by a $\Gamma$-centered effective Hamiltonian. In OsC, the inverted bands at $\Gamma$ remain central to the topological character, but the CBM at K lies below the CBM at $\Gamma$. Consequently, K-valley states may contribute to the electron-doped low-energy response, whereas for Fermi levels inside the gap or near the valence-band edge, the relevant states remain mainly near $\Gamma$.

In the next section, we construct an effective Hamiltonian around $\Gamma$ based on the orbital characters and irreps of the relevant bands. The orbital-resolved band structures are presented in ~\ref{app:orbital}. The irreps at $\Gamma$ were obtained using the irrep subprogram and are listed in Tab.~\ref{tab:irreps_gamma}. The corresponding character and product tables are
given in \ref{app:symmetry}.

Finally, Fig.~\ref{fig:band_structure} shows that a gap is already present at K even without SOC. The relatively flat bands near K may enhance interaction effects, but deriving a K-valley effective Hamiltonian is beyond the scope of this work.

\begin{table}[t]
	\centering
	\caption{Energy positions of the conduction-band minimum (CBM) and valence-band maximum (VBM) of OsC and RuC at the high-symmetry points $\Gamma$, M, and K. Energies are given in meV relative to the Fermi level.}
	\label{table:energy_value}
	\renewcommand{\arraystretch}{0.85}
	\setlength{\tabcolsep}{4.0pt}	
	\begin{tabular}{lccc}
		\hline\hline
		High-symmetry point & $\Gamma$ & M & K \\
		\hline
		CBM (OsC) & 226 & 577 & 110 \\
		VBM (OsC) & -86 & -439 & -1421 \\
		\hline
		CBM (RuC) & 80 & 1411 & 943 \\
		VBM (RuC) & -31 & -342 & -1104 \\
		\hline\hline
	\end{tabular}
\end{table}

\section{Effective Hamiltonian}
\label{sec:effective_hamiltonian}

In this section, we construct a symmetry-constrained $\mathbf{k}\cdot\mathbf{p}$
Hamiltonian near the $\Gamma$ point to describe the low-energy electronic
structure of the monolayer. The basis is chosen as the smallest set of Bloch
states required by the DFT band ordering, orbital composition, and $D_{3h}$
symmetry, including the states closest to the Fermi level and those giving the
dominant symmetry-allowed momentum and SOC. This parent
$8\times8$ Hamiltonian captures the gap formation, leading SOC effects, and the
curvature of the relevant valence and conduction bands. Löwdin partitioning is
then used to downfold the model onto the low-energy subspace, yielding a effective
$4\times4$ Hamiltonian in which the influence of remote bands is absorbed into
renormalized parameters. The remaining parameters are obtained by fitting the
model dispersions to the first-principles band structures.

\subsection{Symmetry constraints on the low-energy \texorpdfstring{$\mathbf{k}\cdot\mathbf{p}$}{k·p} Hamiltonian}
\label{sec:symmetry_kp}

To construct the effective Hamiltonian near the Fermi level, we start from the Bloch theorem~\cite{Bloch1929},
$
	\Phi_{\nu,\mathbf{k}}(\mathbf{r}) = e^{i\mathbf{k}\cdot\mathbf{r}}\,u_{\nu,\mathbf{k}}(\mathbf{r}),
	\label{eq:bloch_theorem}
$
where \(\Phi_{\nu,\mathbf{k}}(\mathbf{r})\) is the Bloch wave function, \(u_{\nu,\mathbf{k}}(\mathbf{r})\) is its cell-periodic part, and \(\nu\) labels the band index.

Since the low-energy states of interest are centered at the \(\Gamma\) point, we expand the Hamiltonian in the vicinity of this point. The eigenvalue problem for the periodic part can then be written as
\begin{equation}
	\left[
	H(\Gamma)+\frac{\hbar^2\mathbf{k}^2}{2m_e}
	+\frac{\hbar}{m_e}\mathbf{k}\cdot\mathbf{p}
	\right]
	u_{\nu,\mathbf{k}}(\mathbf{r})
	=E_\nu(\mathbf{k})\,u_{\nu,\mathbf{k}}(\mathbf{r}),
	\label{eq:kp_eigenvalue}
\end{equation}
where \(m_e\) is the electron mass and \(\mathbf{p}\) is the momentum operator. 

We define the Hamiltonian of the system at the center of the Brillouin zone ($\Gamma$ point) as
$
	H_0 \equiv H(\Gamma).
$
Within the $\mathbf{k}\cdot\mathbf{p}$ framework, the electronic structure in the vicinity of the $\Gamma$ point is obtained by treating the crystal momentum $\mathbf{k}$ as a small perturbation around this high‑symmetry point. Expanding the Hamiltonian in powers of $\mathbf{k}$ leads to additional $\mathbf{k}$‑dependent terms. In particular, the term $\frac{\hbar}{m_e}\mathbf{k}\cdot\mathbf{p}$ introduces momentum‑dependent coupling between different Bloch states and constitutes the central ingredient of the $\mathbf{k}\cdot\mathbf{p}$ method.

For materials containing heavy elements, SOC can significantly influence the electronic band structure. This interaction originates from the coupling between the electron spin and the effective electric field generated by the crystal potential and is incorporated through the SOC Hamiltonian $H_{\mathrm{SOC}}$.

Taking these contributions into account, the eigenvalue problem can be written in the form
\begin{equation}
	\left[
	H_0 + H_{k\cdot p} + H_{\mathrm{SOC}}
	\right]
	u_{\nu,\mathbf{k}}(\mathbf r)
	=
	\epsilon_\nu(\mathbf{k})
	u_{\nu,\mathbf{k}}(\mathbf r),
\end{equation}
where $H_{k\cdot p} = \frac{\hbar}{m_e}\mathbf{k}\cdot\mathbf{p}$.

In the expansion of Eq.~\eqref{eq:kp_eigenvalue}, the term $\frac{\hbar^2 \kv^2}{2m_e}$ also appears, which corresponds to the kinetic energy of a free electron. Since this term represents only a scalar energy shift and does not contribute to interband coupling, it is convenient to absorb it into the definition of the band energy. We therefore introduce the redefined energy
$
	\epsilon_\nu(\mathbf{k})
	=
	E_\nu(\mathbf{k})
	-
	\frac{\hbar^2 \kv^2}{2m_e},
$
where $E_\nu(\mathbf{k})$ is the full band energy. This redefinition removes the trivial free‑electron contribution and yields a more compact form of the effective $\mathbf{k}\cdot\mathbf{p}$ Hamiltonian that captures the relevant interband and spin–orbit interactions.

The basis states at \(\Gamma\) are constructed from the orbital part transforming according to irreps of the little group at \(\Gamma\), combined with spin states. We denote them as
\begin{equation}
	\ket{\psi_{\Gamma_{i\alpha}}^\nu,s}
	=
	u_{\nu,\Gamma}^{\Gamma_{i\alpha}}(\mathbf{r})\otimes\ket{s},
	\label{eq:basis_states_general}
\end{equation}
where \(\Gamma_i\) labels the irrep, \(\alpha\) distinguishes its partner functions when applicable, and \(s=\uparrow,\downarrow\). 
In the present problem, these symmetry-adapted states are identified with the low-energy bands used below to construct the multiband Hamiltonian, labeled as CB, VB, VB--$1$, and VB--$2$. Here, CB denotes the conduction band, VB denotes the valence band, and VB--$i$ denotes the $i$th band below the valence band.

The matrix elements of the \(\kv\cdot \pv\) term between basis states are constrained by crystal symmetry:
\begin{equation}
	\bra{\psi_{\Gamma_{i\alpha}}^\nu,s}
	H_{k\cdot p}
	\ket{\psi_{\Gamma_{i'\alpha'}}^{\nu'},s'}
	=
	\frac{\hbar}{m_e}
	\bra{\psi_{\Gamma_{i\alpha}}^\nu,s}
	\mathbf{k}\cdot\mathbf{p}
	\ket{\psi_{\Gamma_{i'\alpha'}}^{\nu'},s'}.
	\label{eq:kp_matrix_element}
\end{equation}
For the two-dimensional system considered here, it is convenient to use the combinations
\(k_\pm = k_x \pm i k_y\) and \(p_\pm = p_x \pm i p_y\), so that the in-plane coupling can be written as
\begin{equation}
	\mathbf{k}\cdot\mathbf{p}
	=
	\frac{1}{2}\left(k_+p_-+k_-p_+\right).
	\label{eq:kp_pm}
\end{equation}
The symmetry of the monolayer is described by the $D_{3h}$ point group.
The Bloch states at the $\Gamma$ point transform according to the irreps of this group. 

The character table and multiplication rules of the $D_{3h}$ point group are
listed in \ref{app:symmetry}, Tabs.~\ref{tab:D3h_character}
and~\ref{tab:D3h_multiplication}, respectively. They provide the group‑theoretical
basis for determining the symmetry‑allowed $\mathbf{k}\cdot\mathbf{p}$ and
spin–orbit matrix elements.

A matrix elements are non-zero only if the direct product of the irreps contains the totally symmetric
representation:
\begin{equation}
	\Gamma_i^{*}\otimes \Gamma_{\hat{O}}\otimes \Gamma_j \supset A_1' .
	\label{selection_rule}
\end{equation}
The matrix elements appearing in Eq.~\eqref{selection_rule} are constrained by the symmetry of the $D_{3h}$ point group at the $\Gamma$ point. In the
symmetry-adapted basis, a coupling between two states is allowed only when the direct product of the irreps associated with the initial state, the operator component, and the final state contains the totally symmetric representation $A_1'$. This condition acts as the symmetry selection rule for both the $\mathbf{k}\cdot\mathbf{p}$ momentum operators and the SOC terms. As a result, only symmetry-allowed couplings between the relevant $\Gamma$-point states are retained, while all other matrix elements vanish identically. These symmetry constraints determine the structure of the parent $8\times 8$ Hamiltonian constructed below, whose reduced form is used to derive the effective low-energy model. Further details, including the relevant character table, multiplication rules, and allowed matrix elements, are provided in \ref{app:symmetry}.

The spin-orbit interaction is included through \(H_{\mathrm{SOC}}\), whose matrix elements are likewise restricted by symmetry. In terms of the orbital and spin angular momentum operators, the relevant couplings involve \(L_zS_z\) and the ladder-operator combinations \(L_\pm S_\mp\), with \(L_\pm=L_x\pm iL_y\) and \(S_\pm=S_x\pm iS_y\). The corresponding matrix elements are therefore determined by the symmetry properties of the basis states and the allowed operator products.

In practice, the nonzero matrix elements retained in the effective Hamiltonian are determined by combining these symmetry constraints with fits to the \textit{ab initio} band structure in the low-energy window around \(\Gamma\), where the target bands are well separated from higher-energy states. This supports the use of the effective four-band model in the energy range of interest.

\subsection{Eight-band Hamiltonian at the $\Gamma$ point}
\label{sec:eight_band_hamiltonian}

\begin{table}[t]
	\centering
    \caption{irreps of the relevant bands at the $\Gamma$ point.}
	\label{tab:irreps_gamma}
	\renewcommand{\arraystretch}{0.85}
	\setlength{\tabcolsep}{4.0pt}	
	\begin{tabular}{cc}
		\hline\hline
		Band index & Irrep (degeneracy)\\
		\hline
		VB--2      & $A''_2$ (1)\\
		VB--1      & $A'_1$ (1)\\
		VB, CB    & $E''$ (2)\\
		\hline\hline
	\end{tabular}
\end{table}

We now construct an $8\times8$ Hamiltonian at the $\Gamma$ point. We first select four orbital basis states associated with the CB, VB, VB--1, and VB--2 bands. Here, CB denotes the conduction band, VB denotes the valence band, and VB--$i$ denotes the $i$th band below the valence band. Including spin, these four orbital states form an eight-dimensional basis. This multiband Hamiltonian is used as the starting point for deriving the final $4\times4$ low-energy effective Hamiltonian through Löwdin partitioning.

At the $\Gamma$ point, the little group is $D_{3h}$, which coincides with the point group of the crystal. The irreps and degeneracies of the relevant bands are listed in Tab.~\ref{tab:irreps_gamma}. As shown by the orbital-resolved band structures in \ref{app:orbital}, the CB and 
VB states at the $\Gamma$ point are predominantly derived from the 
$d_{xz}$ and $d_{yz}$ orbitals of the transition-metal atom 
$\mathrm{X}$ ($\mathrm{X}=\mathrm{Ru}$ or $\mathrm{Os}$). These states span 
the two-dimensional $E''$ irreducible representation of the $D_{3h}$ point 
group. A convenient complex basis for this subspace is chosen as
$
	\psi_{\pm} \propto d_{xz}(\mathrm{X}) \pm i d_{yz}(\mathrm{X}),
$
which transforms as the $m=\pm1$ components of the $l=2$ spherical 
harmonics, $Y_2^{\pm1}(\theta,\phi)$, up to a convention-dependent phase.

The nonzero matrix elements of the $\kv\cdot\pv$ Hamiltonian, $H_{\kv\cdot\pv}$, and the SOC Hamiltonian, $H_{\mathrm{SOC}}$, are determined using the symmetry selection rules described in Sec.~\ref{sec:symmetry_kp} (Eq.~\eqref{selection_rule}), together with the character and multiplication tables of the $D_{3h}$ point group, as summarized in \ref{app:symmetry}. In addition to the irreps products, the transformation properties of the basis functions and operators under the symmetry operations must be considered explicitly. For example, under the threefold rotation $C_3$, one has
$
	C_3 p_\pm C_3^\dagger=e^{\pm 2i\pi/3}p_\pm,
	C_3 L_\pm C_3^\dagger=e^{\pm 2i\pi/3}L_\pm,
	C_3\psi_\pm=e^{\pm 2i\pi/3}\psi_\pm .
$
These transformation properties determine the allowed couplings involving $k_\pm$, $L_\pm$, $S_\pm$, and $S_z$ in $H_{\kv\cdot\pv}$ and $H_{\mathrm{SOC}}$.

Further constraints are obtained from the mirror operation. Since the basis functions are proportional to spherical harmonics, the mirror operation can be chosen such that
$
	\sigma_y \psi(\theta,\phi)=\psi^*(\theta,\phi).
$
Moreover,
$
	\sigma_y L_z \sigma_y^{-1}=-L_z,
	\sigma_y L_\pm \sigma_y^{-1}=-L_\mp,
	\sigma_y p_\pm \sigma_y^{-1}=p_\mp .
$
With an appropriate phase convention for the basis states, these relations imply that the allowed $\kv\cdot\pv$ matrix elements are purely imaginary, whereas the allowed SOC matrix elements can be chosen to be real.

We denote the relevant effective SOC parameters by $\Delta_1$ and $\Delta_2$. Here, the SOC strength and the relevant spin matrix elements are absorbed into these effective parameters. With the chosen phase convention, they are defined through
$
	\bra{\psi_{\mathrm{cb}}}L_z\ket{\psi_{\mathrm{cb}}}
	=
	-\bra{\psi_{\mathrm{vb}}}L_z\ket{\psi_{\mathrm{vb}}}
	\equiv
	-\Delta_1,
$
and
$
	\bra{\psi_{\mathrm{vb}-1}}L_+\ket{\psi_{\mathrm{cb}}}
	=
	\bra{\psi_{\mathrm{vb}-1}}L_-\ket{\psi_{\mathrm{vb}}}
	\equiv
	\Delta_2 .
$
Similarly, the allowed $\kv\cdot\pv$ matrix elements define the parameters $\gamma_1$ and $\gamma_2$. In particular, the coupling between the CB and VB states is characterized by
$
	\bra{\psi_{\mathrm{cb}}}p_-\ket{\psi_{\mathrm{vb}}}\equiv i\gamma_1,
$
while the coupling involving the VB--2 state is defined through
$
	\bra{\psi_{\mathrm{cb}}}p_+\ket{\psi_{\mathrm{vb}-2}}\equiv i\gamma_2 .
$
The symmetry-allowed momentum matrix elements are summarized in Tab.~\ref{tab:k.p-ham}, while the spin--orbit matrix elements are listed in Tab.~\ref{tab:l.s-ham}.

In the spinful basis
$
\{
\psi_{\mathrm{cb}}\uparrow,
\psi_{\mathrm{cb}}\downarrow,
\psi_{\mathrm{vb}}\uparrow,
\psi_{\mathrm{vb}}\downarrow,
\psi_{\mathrm{vb}-1}\uparrow,
\psi_{\mathrm{vb}-1}\downarrow,
\psi_{\mathrm{vb}-2}\uparrow,
\psi_{\mathrm{vb}-2}\downarrow
\},
$
the $8\times8$ Hamiltonian reads
{\small
	\begin{gather}
		\nonumber
		\mathbf{H}_{8\times8}(\kv)=\\
		\arraycolsep=2pt
		\renewcommand{\arraystretch}{0.95}
		\left(
		\begin{array}{cccccccc}
			\epsilon_{\mathrm{cb}}-\Delta_{1} & 0 & i\gamma_1k_- & 0 & 0 & \Delta_2 & i\gamma_2k_+ & 0 \\
			0 & \epsilon_{\mathrm{cb}}+\Delta _1 & 0 & i\gamma _1k_- & 0 & 0 & 0 & i\gamma_2k_+  \\
			-i\gamma_1 k_+ & 0 & \epsilon_{\mathrm{vb}}+\Delta_1 & 0 & 0 & 0 & -i\gamma_2k_- & 0 \\
			0 & -i\gamma_1 k_+ & 0 & \epsilon_{\mathrm{vb}}-\Delta_1 & \Delta_2 & 0 & 0 & -i\gamma_2k_- \\
			0 & 0 & 0 & \Delta_2 & \epsilon_{\mathrm{vb}-1} & 0 & 0 & 0  \\
			\Delta_2 & 0 & 0 & 0 & 0  & \epsilon_{\mathrm{vb}-1} & 0 & 0 \\
			-i\gamma_2k_- & 0 & i\gamma_2k_+ & 0 & 0  & 0 & \epsilon_{\mathrm{vb}-2} & 0 \\
			0 & -i\gamma_2k_- & 0 & i\gamma_2k_+ & 0  & 0 & 0 & \epsilon_{\mathrm{vb}-2}
		\end{array}
		\right).
		\label{eq:H8x8}
	\end{gather}
}

The parameters $\epsilon_{\mathrm{cb}}$, $\epsilon_{\mathrm{vb}}$, $\epsilon_{\mathrm{vb}-1}$, and $\epsilon_{\mathrm{vb}-2}$ are obtained directly from the band energies at the $\Gamma$ point in the absence of SOC. The remaining parameters, including $\Delta_1$, $\Delta_2$, $\gamma_1$, and $\gamma_2$, are determined by fitting the Hamiltonian to the spin-orbit-coupled \textit{ab initio} band structure.

\subsection{Low-energy four-band Hamiltonian and parameter fitting} 
\label{sec:four_band_hamiltonian}

Based on the symmetry-constrained eight--bands model derived above, we construct an effective 4 $\times$ 4 Hamiltonian for the low-energy bands near the $\Gamma$ point, where the band inversion occurs. The parent 8 $\times$ 8 Hamiltonian contains the CB, VB, VB--1, and VB--2 states at the $\Gamma$ point, together with spin. The on-site energies of these basis states, $\epsilon_{\mathrm{cb}}$, $\epsilon_{\mathrm{vb}}$, $\epsilon_{\mathrm{vb}-1}$, and $\epsilon_{\mathrm{vb}-2}$, are obtained from the non-SOC DFT band structure at $\Gamma$. The symmetry-allowed coupling parameters $\Delta_1$, $\Delta_2$, $\gamma_1$, and $\gamma_2$ are then determined by fitting the model to the SOC first-principles bands in the vicinity of $\Gamma$.

The reduction from the 8 $\times$ 8 Hamiltonian to the four-band model is carried out using Löwdin partitioning. Since RuC and OsC contain heavy transition-metal elements, SOC is not treated as a weak perturbation. Instead, it is incorporated explicitly in the full eight-band Hamiltonian before projection. The downfolding procedure is therefore used only to eliminate remote bands that are well separated in energy from the low-energy subspace. The resulting 4 $\times$ 4 model is valid in the immediate vicinity of $\Gamma$, where the target bands are isolated and govern the band-inversion physics.

We first diagonalize the $\kv=0$ Hamiltonian including SOC. This defines the low-energy basis states as
\begin{equation}
	\begin{split}
		\ket{\Phi_1} &=
		\cos\theta\ket{\psi_{\mathrm{cb}},\uparrow}
		-
		\sin\theta\ket{\psi_{\mathrm{vb}-1},\downarrow},\\
		\ket{\Phi_2} &=
		\ket{\psi_{\mathrm{cb}},\downarrow},\\
		\ket{\Phi_3} &=
		\ket{\psi_{\mathrm{vb}},\uparrow},\\
		\ket{\Phi_4} &=
		\cos\theta\ket{\psi_{\mathrm{vb}},\downarrow}
		-
		\sin\theta\ket{\psi_{\mathrm{vb}-1},\uparrow}.
	\end{split}
	\label{eq:four_band_basis}
\end{equation}
Here, $\theta$ is the SOC-induced mixing angle obtained by diagonalizing
the $\kv=0$ part of the 8 $\times$ 8 Hamiltonian. The orthogonal
combinations generated by the same diagonalization belong to the
remote subspace and are eliminated in the subsequent Löwdin projection.

At $\kv=0$, the SOC Hamiltonian
$
	H_{\mathrm{SOC}}=\lambda\,\mathbf{L}\cdot\mathbf{S}
$
couples only specific pairs of states through the spin ladder operators
$S_\pm$ contained in $\mathbf{L}\cdot\mathbf{S}$.
Consistent with the symmetry selection rules discussed in~\ref{app:symmetry}, the allowed SOC matrix elements connect
$(\psi_{\mathrm{cb}},\uparrow)$ with $(\psi_{\mathrm{vb}-1},\downarrow)$
and $(\psi_{\mathrm{vb}},\downarrow)$ with $(\psi_{\mathrm{vb}-1},\uparrow)$,
while the states $\ket{\psi_{\mathrm{cb}},\downarrow}$ and
$\ket{\psi_{\mathrm{vb}},\uparrow}$ remain uncoupled at $\kv=0$.
As a result, the Hamiltonian separates into independent $2\times2$
blocks, and the SOC-induced mixing appearing in
Eq.~\eqref{eq:four_band_basis} arises from the diagonalization
of these blocks.

After transforming the full 8 $\times$ 8 Hamiltonian into the SOC-mixed low-energy subspace, the remote states are eliminated by Löwdin partitioning, retaining terms up to second order in $\kv$. In the basis
$
\{
\ket{\Phi_1},
\ket{\Phi_2},
\ket{\Phi_3},
\ket{\Phi_4}
\},	
$
the resulting $4\times4$ effective Hamiltonian reads
{\setlength{\arraycolsep}{3pt}
	\renewcommand{\arraystretch}{0.95}
	\begin{equation}
		\label{eq:4x4ham}
		H_{4\times4}(\kv)=
		\left(
		\begin{array}{cccc}
			\mathcal{H}_1(\kv) & 0 & \mathcal{H}_3(\kv) & 0 \\
			0 & \mathcal{H}_2(\kv) & 0 & \mathcal{H}_3(\kv) \\
			\mathcal{H}_3^*(\kv) & 0 & \mathcal{H}_2(\kv) & 0 \\
			0 & \mathcal{H}_3^*(\kv) & 0 & \mathcal{H}_1(\kv)
		\end{array}
		\right),
	\end{equation}
}
with
\begin{equation}
	\mathcal{H}_{1,2}(\kv)=A_{1,2}+B_{1,2}k^2,
	\label{eq:H12}
\end{equation}
and
\begin{equation}
	\mathcal{H}_{3}(\kv)=iNk_-+B_3k_+^2.
	\label{eq:H3}
\end{equation}
Here, $k^2=k_x^2+k_y^2$. The diagonal terms $\mathcal{H}_{1,2}$ describe the leading parabolic dispersions of the two low-energy branches, whereas $\mathcal{H}_3$ accounts for their momentum-dependent hybridization. The terms proportional to $N$ and $B_3$ represent the leading symmetry-allowed interband couplings retained up to second order in $\kv$. The parameters $A_{1,2}$, $B_{1,2}$, $N$, and $B_3$ are effective, renormalized quantities resulting from the projection and are obtained by fitting the 4 $\times$ 4 Hamiltonian to the spin-orbit-coupled first-principles band structure near $\Gamma$ point.

For both OsC and RuC, the fitted value of the linear coefficient $N$ is negligible,
consistent with $\gamma_1\simeq 0$ in Tab.~\ref{table:fitting_value}. Consequently,
although the linear term $iNk_-$ is allowed by symmetry, the dominant interband
coupling is provided by the quadratic term $B_3k_+^2$. This term is not introduced
ad hoc; instead, it arises from symmetry-allowed coupling to remote bands through
the Löwdin downfolding procedure. It governs the orbital mixing away from $\Gamma$
and is essential for reproducing the finite-$k$ dispersion obtained from
first-principles calculations.

The structure of Eq.~\eqref{eq:4x4ham} is reminiscent of the BHZ model, in the
sense that it consists of two two-band sectors related by time-reversal symmetry.
However, the present Hamiltonian exhibits a material-specific feature: the dominant
off-diagonal hybridization is quadratic, rather than linear, in momentum. This
behavior reflects the symmetry character of the low-energy states in RuC and OsC.
Although these quadratic terms vanish at $\Gamma$ and therefore do not change the
band inversion at the zone center, they strongly affect the dispersion at finite-$k$ and are required for a quantitative description of the low-energy
bands.

The parameters of the effective $\mathbf{k}\cdot\mathbf{p}$ Hamiltonian were obtained by
fitting its eigenvalues to the spin--orbit-coupled DFT bands in the vicinity
of the $\Gamma$ point, following the $\mathbf{k}\cdot\mathbf{p}$ fitting scheme of
Ref.~\cite{oliver}. The optimized quantity was a weighted absolute
band-energy mismatch,
\begin{equation}
	\sum_{j,\kv}
	w_{j,\kv}
	\left|
	E_j^{\mathrm{kp}}(\kv;P)
	-
	E_j^{\mathrm{DFT}}(\kv)
	\right| ,
	\label{eq:fitting_function}
\end{equation}
where $P$ denotes the set of fitted Hamiltonian parameters listed in
Tab.~\ref{table:fitting_value}. The weights $w_{j,\kv}$ encode the relative
importance assigned to different bands and $k$ points, with larger weights
given to the low-energy bands near the Fermi level and to states closer to
$\Gamma$. The fitting range was limited by the validity of the Löwdin
partitioning: the $k$-dependent terms coupling the low-energy subspace to
remote bands must remain small compared with the corresponding energy
separation. We therefore restricted the fitting to
$|\kv|<0.1~\text{\AA}^{-1}$ for OsC and $|\kv|<0.05~\text{\AA}^{-1}$ for RuC,
where the low-energy bands remain well separated from remote bands. These
intervals define the range of validity of the fitted low-energy $\mathbf{k}\cdot\mathbf{p}$
model. The fitted bands are shown by the red dashed lines in
Fig.~\ref{fig:band_structure}, and the corresponding parameters are listed in
Tab.~\ref{table:fitting_value}.

\begin{table}[ht]
	\centering
	\caption{
		Parameters of the 8 $\times$ 8 $\mathbf{k}\!\cdot\!\mathbf{p}$ Hamiltonian and
		the corresponding coefficients of the effective 4 $\times$ 4 Hamiltonian
		for monolayer OsC and RuC. Parameters $A_3$ and $A_4$ originate from the
		parent 8 $\times$ 8 Hamiltonian [Eq.~\eqref{eq:A_params}] and are retained for completeness. After
		the Löwdin partitioning, only $A_1$ and $A_2$ appear explicitly in the
		effective Hamiltonian [Eq.~\eqref{eq:4x4ham}].
	}
	\label{table:fitting_value}
	\renewcommand{\arraystretch}{0.9}
	\begin{tabular}{l c c @{\hspace{1.5cm}} l c c}
		\toprule
		\multicolumn{3}{c}{8$\times$8 Hamiltonian} &
		\multicolumn{3}{c}{4$\times$4 effective Hamiltonian} \\
		\cmidrule(r){1-3}\cmidrule(l){4-6}
		Parameter & OsC & RuC & Parameter & OsC & RuC \\
		\midrule
		$\epsilon_{\mathrm{cb}}$ (eV)   & 0.000   & 0.000   & $A_1$ (eV)          & -0.085   & -0.031 \\
		$\epsilon_{\mathrm{vb}}$ (eV)   & 0.000   & 0.000   & $A_2$ (eV)          & 0.226    & 0.080 \\
		$\epsilon_{\mathrm{vb}-1}$ (eV) & -1.259  & -0.628  & $A_3$ (eV)          & -1.400   & -0.677 \\
		$\epsilon_{\mathrm{vb}-2}$ (eV) & -2.198  & -1.531  & $A_4$ (eV)          & -2.198   & -1.531 \\
		$\Delta_1$ (eV)                 & 0.226   & 0.080   & $B_1$ (eV\,\AA$^2$) & 114.907  & 134.675 \\
		$\Delta_2$ (eV)                 & 0.407   & 0.171   & $B_2$ (eV\,\AA$^2$) & 112.287  & 135.654 \\
		$\cos\theta$                    & 0.945   & 0.961   & $B_3$ (eV\,\AA$^2$) & -110.038 & -131.438 \\
		$\gamma_1$ (eV\,\AA)            & 0.000   & 0.000   &                     &          &         \\
		$\gamma_2$ (eV\,\AA)            & -16.215 & -14.574 &                     &          &         \\
		\bottomrule
	\end{tabular}
\end{table}

Since the fitted values give $N\simeq 0$ (Eq.~\eqref{eq:A_params}), the linear coupling term can be neglected. 
In the reordered basis 
$\{\ket{\Phi_1},\ket{\Phi_3},\ket{\Phi_2},\ket{\Phi_4}\}$, 
Eq.~\eqref{eq:4x4ham} reduces to the block-diagonal form
\begin{equation}
	H_{4\times4}(\kv)=
	\begin{pmatrix}
		H_{+}(\kv) & 0 \\
		0 & H_{-}(\kv)
	\end{pmatrix},
	\label{eq:block_diagonal}
\end{equation}
with
\begin{gather}
	H_{+}(\kv)=
	\begin{pmatrix}
		A_1+B_1k^2 & B_3k_+^2 \\
		B_3k_-^2 & A_2+B_2k^2
	\end{pmatrix},
	\nonumber\\
	H_{-}(\kv)=
	\begin{pmatrix}
		A_2+B_2k^2 & B_3k_+^2 \\
		B_3k_-^2 & A_1+B_1k^2
	\end{pmatrix}.
	\label{eq:2x2_blocks}
\end{gather}
Here, $H_{+}$ and $H_{-}$ act in the subspaces 
$\{\ket{\Phi_1},\ket{\Phi_3}\}$ and 
$\{\ket{\Phi_2},\ket{\Phi_4}\}$, respectively, and form a time-reversal pair. 
Because the basis states include SOC-induced mixing, the labels $+$ and $-$ do not denote pure spin sectors.

The block $H_{+}(\mathbf{k})$ can be written in the Pauli-matrix form in the
$\{\ket{\Phi_1},\ket{\Phi_3}\}$ subspace as
\begin{equation}
	H_{+}(\mathbf{k})=
	d_0(\mathbf{k})\sigma_0+
	d_1(\mathbf{k})\sigma_x+
	d_2(\mathbf{k})\sigma_y+
	d_3(\mathbf{k})\sigma_z ,
	\label{eq:pauli_expansion}
\end{equation}
where $\sigma_0$ denotes the identity matrix and $\sigma_{x,y,z}$ are the
Pauli matrices acting in this two-state subspace. The coefficients are
\begin{equation}
	\begin{aligned}
		d_0(\mathbf{k}) &=
		\frac{A_1+A_2}{2}
		+
		\frac{B_1+B_2}{2}k^2, \\
		d_1(\mathbf{k}) &=
		B_3(k_x^2-k_y^2), \\
		d_2(\mathbf{k}) &=
		-2B_3 k_x k_y, \\
		d_3(\mathbf{k}) &=
		\frac{A_1-A_2}{2} + \frac{B_1-B_2}{2}k^2 .
	\end{aligned}
	\label{eq:d_coefficients}
\end{equation}

The corresponding eigenvalues are
\begin{equation}
	E_{\pm}(\kv)=
	d_0(\kv)\pm
	\sqrt{
		d_1^2(\mathbf{k})+
		d_2^2(\mathbf{k})+
		d_3^2(\mathbf{k})
	} .
	\label{eq:pauli_eigenvalues}
\end{equation}
Equivalently, using
$d_1^2(\mathbf{k})+d_2^2(\mathbf{k})=B_3^2 k^4$, the spectrum can be written as
$
	E_{\pm}(k)=
	d_0(k)\pm
	\sqrt{
		d_3^2(k)+B_3^2 k^4
	} .
	\label{eq:pauli_eigenvalues_simplified}
$

Because the full $4\times4$ Hamiltonian consists of two
time-reversal-related blocks, each energy branch is doubly degenerate.
In this representation, $d_0(\kv)$ describes the average dispersion of the
two bands, while $d_3(\kv)$ acts as an effective mass term that determines
their relative ordering. In particular, $d_3(0)=(A_1-A_2)/2$ fixes the
band configuration at the $\Gamma$ point, and its sign distinguishes the
normal and inverted regimes. The functions $d_1(\mathbf{k})$ and
$d_2(\mathbf{k})$ encode the symmetry-allowed interband hybridization that
couples the two states away from the zone center.

Additional insight into the structure of the coupling is obtained by
writing $k_\pm = k e^{\pm i\phi}$. The quadratic term $B_3 k_\pm^2$
therefore carries the angular phase $e^{\pm 2 i\phi}$, implying that the
in-plane pseudospin components $(d_1,d_2)$ wind twice as $\mathbf{k}$
encircles the $\Gamma$ point. This double-winding behavior directly
reflects the quadratic nature of the interband coupling and contrasts with
conventional BHZ-type models, where the dominant hybridization is linear
in momentum. The analytical spectrum obtained from this representation
accurately reproduces the low-energy bands shown in
Fig.~\ref{fig:band_structure}.

Notably, the quadratic coupling $B_3 k_\pm^2$ implies that the in-plane
pseudospin components behave as $(d_1,d_2)\propto k^2(\cos2\phi,-\sin2\phi)$.
Consequently, the pseudospin winds twice as $\mathbf{k}$ encircles the
$\Gamma$ point, corresponding to an effective angular-momentum channel
$l=2$. This contrasts with conventional BHZ-type model, where the
hybridization is linear in momentum and produces a single winding
($l=1$).

For $N\simeq0$, the two time-reversed blocks have identical eigenvalues,
\begin{equation}
	E_{\pm}(\kv)
	=
	\frac{
		A_1+A_2+(B_1+B_2)k^2
	}{2}
	\pm
	\sqrt{
		\left[
		\frac{
			A_1-A_2+(B_1-B_2)k^2
		}{2}
		\right]^2
		+
		B_3^2 k^4
	}.
	\label{eq:four_band_eigenvalues}
\end{equation}
Each branch is twofold degenerate because the two blocks form a time-reversal pair. 

The low-energy Hamiltonian derived here is based on the static-lattice approximation, and electron--phonon coupling is not included explicitly. Such interactions may renormalize the band-edge energies, the topological gap, and the fitted effective parameters at finite temperature, and may also contribute to quasiparticle broadening. As long as the relevant crystal symmetries are preserved, however, they do not change the symmetry-allowed form of the low-energy Hamiltonian. Their principal effect can therefore be viewed as a temperature-dependent renormalization of the model parameters. A quantitative first-principles treatment of electron--phonon coupling, including its impact on carrier lifetimes, mobility, transport properties, and possible temperature-dependent gap renormalization, lies beyond the scope of the present work and will be considered in future studies.

\subsection{Apparent masses and validity of the $\Gamma$-centered $\mathbf{k}\cdot\mathbf{p}$ description}

Effective masses reported from DFT band structures are often obtained by fitting the dispersion over a finite momentum interval around a band extremum. Such fitted values represent apparent masses and may differ from the strictly local curvature mass when the underlying band is nonparabolic~\cite{Whalley2019,Cockton2026}. Within $\mathbf{k}\cdot\mathbf{p}$ theory, the curvature of a given band is renormalized by interband coupling, including coupling to remote bands~\cite{LuttingerKohn1955,Foreman1997}. The Kane model provides the classical example of this mechanism, showing that interband coupling can produce a nonparabolic conduction-band dispersion and an energy-dependent electron effective mass~\cite{Kane1957}. The local curvature mass itself can be rigorously defined through the Hessian of the electronic dispersion and evaluated directly within density-functional perturbation theory~\cite{Laflamme2016}.

To make a consistent comparison with the DFT results, we therefore extract apparent masses from the analytic four-band $\mathbf{k}\cdot\mathbf{p}$ dispersion in Eq.~\eqref{eq:four_band_eigenvalues} using the same fitting procedure. In the local limit, $k\rightarrow 0$, the corresponding curvature masses are given by $m^*_{1,2}=\hbar^2/(2B_{1,2})$. Away from this limit, however, interband hybridization in the 4 $\times$ 4 Hamiltonian generates higher-order terms in $\kv$, so that the fitted apparent mass becomes window-dependent.

Tab.~\ref{tab:kp_vs_dft_bestmatch} compares the best-matching $\mathbf{k}\cdot\mathbf{p}$ apparent masses with the DFT apparent masses for OsC and RuC. For the valence bands of both materials, the agreement is excellent: the relative deviations are below $1\%$. This confirms that the 4 $\times$ 4 Hamiltonian captures the valence-band curvature reliably within the considered $\Gamma$-centered subspace. By contrast, the conduction-band apparent masses obtained from the four-band model remain substantially smaller than the corresponding DFT values, even when the fitting window is optimized.

\begin{table}[t]
	\centering
	\caption{Best-matching $\mathbf{k}\cdot\mathbf{p}$ apparent masses compared with DFT apparent masses for OsC and RuC. The optimal fitting window $k_{\max}$ minimizes the deviation between the two fitted values.}
	\label{tab:kp_vs_dft_bestmatch}
	\small
	\renewcommand{\arraystretch}{0.9}
	\setlength{\aboverulesep}{0pt}
	\setlength{\belowrulesep}{0pt}
	\begin{tabular}{lccccc}
		\toprule
		Material & Band & $k_{\max}$ ($\mathrm{\AA^{-1}}$) & $m^*_{k\cdot p,\mathrm{app}}/m_0$ & $m^*_{\mathrm{DFT,app}}/m_0$ & Error (\%) \\
		\midrule
		OsC & VB & 0.116 & 0.2843 & 0.2851 & 0.28 \\
		OsC & CB & 0.005 & 0.0337 & 0.0775 & 56.5 \\
		RuC & VB & 0.055 & 0.1922 & 0.1934 & 0.62 \\
		RuC & CB & 0.005 & 0.0274 & 0.1223 & 77.6 \\
		\bottomrule
	\end{tabular}
\end{table}

This difference reflects the limited quantitative accuracy of the present $\Gamma$-centered 4 $\times$ 4 description for the conduction sector. The VBs are relatively well isolated in energy, making the selected 4 $\times$ 4 subspace sufficient to reproduce their curvature. The CBs, however, are more sensitive to coupling with higher-lying remote bands omitted from the present basis; such remote-band contributions are known to renormalize effective masses and band curvatures in multiband $\mathbf{k}\cdot\mathbf{p}$ theory \cite{Foreman1997,Vurgaftman2001}. This limitation is further amplified in OsC, where the actual CBM is located at $K$ rather than at $\Gamma$. Therefore, the present $\mathbf{k}\cdot\mathbf{p}$ Hamiltonian is quantitatively reliable for describing the valence bands near $\Gamma$, whereas an extended model including additional conduction and remote bands would be required for a fully accurate description of the electron effective masses.

\section{Conclusions}
\label{sec:conclusion}

We have constructed a symmetry-constrained low-energy effective Hamiltonian for
monolayer OsC and RuC in their planar hexagonal phases. Building on previous
first-principles studies that identified these compounds as two-dimensional
topological insulators, our work provides, to the best of our knowledge, the
first analytical \(\mathbf{k}\cdot\mathbf{p}\) description derived from explicit
symmetry considerations at the \(\Gamma\) point. The model establishes a compact
connection between the DFT band structures and the microscopic origin of the
quantum spin Hall phase in these systems.

Starting from an 8 $\times$ 8 spin-orbit-coupled parent Hamiltonian, we used
Löwdin partitioning to derive a \(4\times4\) effective model valid near
\(\Gamma\). Because SOC is included already in the parent Hamiltonian, it is not
treated as a small perturbation in the reduction. The resulting model retains
the relevant band-edge states and captures the SOC-induced band inversion
responsible for the topological phase.

A central feature of the effective Hamiltonian is the presence of
symmetry-allowed quadratic off-diagonal momentum-dependent terms. Unlike the
linear hybridization in the minimal BHZ model, these terms follow from the
irreps of the RuC and OsC basis states at \(\Gamma\) point. They
are essential for reproducing the DFT dispersions near the zone center, together
with the fitted diagonal quadratic terms that determine the band curvatures. The
model agrees well with the first-principles bands within the fitting windows
\(|\kv|<0.1~\mathrm{\AA}^{-1}\) for OsC and
\(|\kv|<0.05~\mathrm{\AA}^{-1}\) for RuC. A comparison of finite-window apparent effective masses extracted from the
analytic dispersion and from the DFT bands further confirms the validity of the
model for the valence bands, while larger deviations for the conduction bands
indicate the influence of remote-band effects beyond the present
$\Gamma$-centered description.

For RuC, the effective Hamiltonian directly describes the low-energy bands near
the Fermi level. For OsC, where the calculated conduction minimum at K lies
below that at \(\Gamma\), the present \(\Gamma\)-centered model is most directly
applicable to the gap and valence-edge physics, while electron-doped regimes may
require an additional K-valley description. More generally, the symmetry-based
\(4\times4\) Hamiltonian provides a tractable starting point for studying
strain, electric fields, disorder, finite-size effects, edge-state transport,
and topological phase transitions in OsC- and RuC-based nanostructures.

\section*{Acknowledgments}
\appendix
\section{Orbital-resolved band structure and basis functions}
\label{app:orbital}

\begin{figure}[h]
	\centering
	\includegraphics[width=1.05\textwidth]{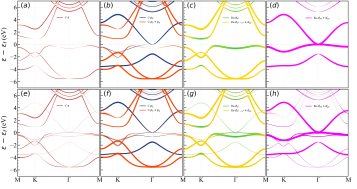}
	\caption{
		Orbital-resolved band structure of monolayer RuC in the hexagonal phase.
		Panels (a)--(d) show the spinless results without SOC, whereas
		panels (e)--(h) show the spinful results with SOC included. The first
		and second columns display the orbital contributions from C atoms, while
		the third and fourth columns display those from Ru atoms. All orbital
		weights are plotted on a common scale; thicker lines indicate larger
		orbital contributions.
	}
	\label{fig:orbital_resolved}
\end{figure}

To construct the effective low-energy Hamiltonian, it is necessary to identify the orbital character of the band-edge states near the $\Gamma$ point. Fig.~\ref{fig:orbital_resolved} shows the orbital-resolved band structures of monolayer RuC without and with SOC. The projections reveal that the CB and VB edge states around the Fermi level are mainly derived from transition-metal $d$ orbitals, whereas the nearby VBs VB--$1$ and VB--$2$ have significant contributions from Ru $d_{z^2}$, C $s$, and C $p_z$ orbitals.

In particular, the CB and VB states at $\Gamma$ are dominated by the Ru $d_{xz}$ and $d_{yz}$ orbitals, which form a twofold degenerate manifold in the absence of SOC. The VB--$1$ band mainly originates from a hybridization of Ru $d_{z^2}$ and C $s$ orbitals, while the VB--$2$ band is primarily composed of the C $p_z$ orbital.

Based on this orbital composition, the minimal basis used to construct the effective $\mathbf{k}\cdot\mathbf{p}$ Hamiltonian consists of four states
$
	|\psi_{cb}\rangle \sim d_{xz}-i d_{yz}, 
	|\psi_{vb}\rangle \sim d_{xz}+i d_{yz}, 
	|\psi_{vb-1}\rangle \sim s+d_{z^2}, 
	|\psi_{vb-2}\rangle \sim p_z .
$

The first two states form the band-edge doublet originating from the $d_{xz}$ and $d_{yz}$ orbitals of Ru, while the remaining two states correspond to the nearby VBs that provide additional coupling channels in the effective Hamiltonian.

Although the effective model discussed in the main text is formulated in terms of four bands, the orbital mixing induced by SOC indicates that a larger parent Hamiltonian is required for a systematic derivation. In practice, we first construct an 8 $\times$ 8 parent Hamiltonian including orbital and spin degrees of freedom, which is then reduced to an effective 4 $\times$ 4 model using Löwdin partitioning.

The orbital analysis presented here provides the physical basis for the symmetry classification of these states discussed in~\ref{app:symmetry}.

\section{Symmetry analysis at the $\Gamma$ point}
\label{app:symmetry}

The low-energy electronic states of the RuC and OsC monolayers at the
$\Gamma$ point are classified according to the irreps of the $D_{3h}$ point group.
The corresponding character and multiplication tables are listed in
Tabs.~\ref{tab:D3h_character} and~\ref{tab:D3h_multiplication},
adapted from Ref.~\cite{altmann1994}. These symmetry relations are used
to determine the symmetry-allowed terms in the effective multiband
Hamiltonian.

\begin{table}[h]
	\centering
	\caption{Character table of the $D_{3h}$ point group.}
	\label{tab:D3h_character}
	\renewcommand{\arraystretch}{0.75}
	\setlength{\tabcolsep}{4.0pt}
	\begin{tabular}{@{}lccccccc@{}}
		\toprule
		Irrep & $E$ & $2C_3$ & $3C_2'$ & $\sigma_h$ & $2S_3$ & $3\sigma_v$ & Basis \\
		\midrule
		$A_1'$  & 1 & 1  & 1  & 1  & 1  & 1  & -- \\
		$A_2'$  & 1 & 1  & -1 & 1  & 1  & -1 & $R_z$ \\
		$E'$    & 2 & -1 & 0  & 2  & -1 & 0  & $(x,y)$ \\
		$A_1''$ & 1 & 1  & 1  & -1 & -1 & -1 & -- \\
		$A_2''$ & 1 & 1  & -1 & -1 & -1 & 1  & $z$ \\
		$E''$   & 2 & -1 & 0  & -2 & 1  & 0  & $(R_x,R_y)$ \\
		\bottomrule
	\end{tabular}
\end{table}

\begin{table}[h]
	\centering
	\caption{Multiplication table of the $D_{3h}$ irreducible representations.}
	\label{tab:D3h_multiplication}
	\renewcommand{\arraystretch}{0.75}
	\setlength{\tabcolsep}{4.0pt}
	\begin{tabular}{@{}c|cccccc@{}}
		\toprule
		$\otimes$ & $A_1'$ & $A_2'$ & $E'$ & $A_1''$ & $A_2''$ & $E''$ \\
		\midrule
		$A_1'$  & $A_1'$ & $A_2'$ & $E'$ & $A_1''$ & $A_2''$ & $E''$ \\
		$A_2'$  & $A_2'$ & $A_1'$ & $E'$ & $A_2''$ & $A_1''$ & $E''$ \\
		$E'$    & $E'$ & $E'$ & $A_1'\!\oplus\!A_2'\!\oplus\!E'$ & $E''$ & $E''$ & $A_1''\!\oplus\!A_2''\!\oplus\!E''$ \\
		$A_1''$ & $A_1''$ & $A_2''$ & $E''$ & $A_1'$ & $A_2'$ & $E'$ \\
		$A_2''$ & $A_2''$ & $A_1''$ & $E''$ & $A_2'$ & $A_1'$ & $E'$ \\
		$E''$   & $E''$ & $E''$ & $A_1''\!\oplus\!A_2''\!\oplus\!E''$ & $E'$ & $E'$ & $A_1'\!\oplus\!A_2'\!\oplus\!E'$ \\
		\bottomrule
	\end{tabular}
\end{table}

The orbital analysis presented in~\ref{app:orbital} shows that
the relevant states near the Fermi level transform as
$
	\psi_{cb},\psi_{vb} \rightarrow E'', 
	\psi_{vb-1} \rightarrow A_1', 
	\psi_{vb-2} \rightarrow A_2''.
$

Including spin, the full basis used in the multiband Hamiltonian is
\begin{equation}
	\left\{
	\psi_{cb},\psi_{vb},\psi_{vb-1},\psi_{vb-2}
	\right\}\otimes
	\left\{\uparrow,\downarrow\right\}.
\end{equation}

The allowed matrix elements follow from the group-theoretical
selection rule
\begin{equation}
	\Gamma_i^{*}\otimes \Gamma_{\hat{O}}\otimes \Gamma_j
	\supset A_1'.
	\label{eq:selection_rule}
\end{equation}
Operators that do not satisfy this condition lead to vanishing
matrix elements by symmetry.

An important application of this rule concerns the
$\mathbf{k}\cdot\mathbf{p}$ coupling. The in-plane momentum operators
$(k_x,k_y)$ transform as the $E'$ representation of the $D_{3h}$
point group. Therefore, the momentum-dependent couplings between
the basis states must satisfy Eq.~(\eqref{eq:selection_rule}) with
$\Gamma_{\hat O}=E'$.

Applying this condition shows that symmetry allows linear-in-$k$
couplings between the $E''$ doublet $(\psi_{cb},\psi_{vb})$
and the $A_2''$ state $\psi_{vb-2}$, as well as between the two
$E''$ states themselves. These symmetry-allowed terms give rise to
the coupling parameters $\gamma_1$ and $\gamma_2$ appearing in the
effective $\mathbf{k}\cdot\mathbf{p}$ Hamiltonian of the main text.
Matrix elements that violate the selection rule vanish identically.

\section{$\mathbf{k}\cdot\mathbf{p}$ and spin--orbit matrix elements}
\label{app:kp_soc}

Using the symmetry analysis presented in~\ref{app:symmetry},
we determine the explicit form of the $\mathbf{k}\cdot\mathbf{p}$ and
SOC matrix elements at the $\Gamma$ point.

The in-plane momentum components can be written in circular form as
$
	k_{\pm}=k_x \pm i k_y .
$
The symmetry-allowed $\mathbf{k}\cdot\mathbf{p}$ matrix elements
between the basis states are therefore we have:

\begin{table}[h]
	\centering
	\caption{$\mathbf{k}\cdot\mathbf{p}$ matrix elements at the $\Gamma$ point.}
	\small
	\renewcommand{\arraystretch}{0.75}
	\setlength{\tabcolsep}{4.0pt}
	\begin{tabular}{lcccc}
		\toprule
		& CB ($E''$) & VB ($E''$) & VB$-1$ ($A'_1$) & VB$-2$ ($A''_2$) \\
		\midrule
		CB ($E''$)      & 0 & $i\gamma_1 k_-$ & 0 & $i\gamma_2 k_+$ \\
		VB ($E''$)      & $-i\gamma_1 k_+$ & 0 & 0 & $-i\gamma_2 k_-$ \\
		VB$-1$ ($A'_1$) & 0 & 0 & 0 & 0 \\
		VB$-2$ ($A''_2$)& $-i\gamma_2 k_-$ & $i\gamma_2 k_+$ & 0 & 0 \\
		\bottomrule
	\end{tabular}
	\label{tab:k.p-ham}
\end{table}
Here $\gamma_1$ and $\gamma_2$ describe the strength of the linear momentum couplings allowed by symmetry.

The spin--orbit interaction is described by the atomic Hamiltonian
$
	H_{\mathrm{SOC}}=\lambda\,\mathbf{L}\cdot\mathbf{S}.
$
Applying the same symmetry considerations yields the SOC matrix elements are:

\begin{table}[h]
	\centering
	\caption{Spin--orbit ($L\!\cdot\!S$) matrix elements at the $\Gamma$ point.}
	\small
	\renewcommand{\arraystretch}{0.75}
	\setlength{\tabcolsep}{4.0pt}	
	\begin{tabular}{lcccc}
		\toprule
		& CB ($E''$) & VB ($E''$) & VB$-1$ ($A'_1$) & VB$-2$ ($A''_2$) \\
		\midrule
		CB ($E''$)      & $-\Delta_1 S_z$ & 0 & $\Delta_2 S_+$ & 0 \\
		VB ($E''$)      & 0 & $\Delta_1 S_z$ & $\Delta_2 S_-$ & 0 \\
		VB$-1$ ($A'_1$) & $\Delta_2 S_-$ & $\Delta_2 S_+$ & 0 & 0 \\
		VB$-2$ ($A''_2$)& 0 & 0 & 0 & 0 \\
		\bottomrule
	\end{tabular}
	\label{tab:l.s-ham}
\end{table}

The parameters $\gamma_1$ and $\gamma_2$ determine the strength of
the $\mathbf{k}\cdot\mathbf{p}$ coupling, while $\Delta_1$ and
$\Delta_2$ characterize the SOC-induced mixing between the bands.
These parameters define the structure of the $8\times8$ Hamiltonian
given in Eq.~\eqref{eq:H8x8} of the main text.

\section{\texorpdfstring{$4\times4$}{4x4} effective 
	\texorpdfstring{$\mathbf{k}\cdot\mathbf{p}$}{k.p} Hamiltonian}
\label{app:4by4_kp_hamiltonian}

In this appendix, we derive the 4 $\times$ 4 effective Hamiltonian starting from the
8 $\times$ 8 parent Hamiltonian. We first diagonalize the SOC-containing
Hamiltonian at the $\Gamma$ point, $H_{8\times8}(\mathbf{k}=\mathbf{0})$, and
then apply Löwdin partitioning to eliminate the remote bands. In this procedure,
SOC is included non-perturbatively in the parent Hamiltonian, while the
perturbative expansion is used only to downfold the high-energy states.

\begingroup
\scriptsize
\setlength{\arraycolsep}{2pt}
\renewcommand{\arraystretch}{0.95}
\begin{equation}
	\mathbf{H}_{8\times8}(\mathbf{k}=\mathbf{0})=
	\begin{pmatrix}
		\epsilon_{\mathrm{cb}}-\Delta_1 & 0 & 0 & 0 & 0 & \Delta_2 & 0 & 0 \\
		0 & \epsilon_{\mathrm{cb}}+\Delta_1 & 0 & 0 & 0 & 0 & 0 & 0 \\
		0 & 0 & \epsilon_{\mathrm{vb}}+\Delta_1 & 0 & 0 & 0 & 0 & 0 \\
		0 & 0 & 0 & \epsilon_{\mathrm{vb}}-\Delta_1 & \Delta_2 & 0 & 0 & 0 \\
		0 & 0 & 0 & \Delta_2 & \epsilon_{\mathrm{vb}-1} & 0 & 0 & 0 \\
		\Delta_2 & 0 & 0 & 0 & 0 & \epsilon_{\mathrm{vb}-1} & 0 & 0 \\
		0 & 0 & 0 & 0 & 0 & 0 & \epsilon_{\mathrm{vb}-2} & 0 \\
		0 & 0 & 0 & 0 & 0 & 0 & 0 & \epsilon_{\mathrm{vb}-2}
	\end{pmatrix}.
\end{equation}
\endgroup

Here, the energy reference is chosen such that
$\epsilon_{\mathrm{cb}}=\epsilon_{\mathrm{vb}}=0$, while
$\epsilon_{\mathrm{vb}-1}$ and $\epsilon_{\mathrm{vb}-2}$ are obtained from the
spinless \textit{ab initio} band structure (Tab.~\ref{table:fitting_value}). The Hamiltonian
$H_{8\times8}(\mathbf{k}=\mathbf{0})$ is diagonalized by the unitary
transformation

\begingroup
\scriptsize
\setlength{\arraycolsep}{2.5pt}
\renewcommand{\arraystretch}{0.95}
\begin{equation}
	U=
	\left(
	\begin{array}{cccccccc}
		\cos\theta & 0 & 0 & 0 & 0 & \sin\theta & 0 & 0 \\
		0 & 1 & 0 & 0 & 0 & 0 & 0 & 0 \\
		0 & 0 & 1 & 0 & 0 & 0 & 0 & 0 \\
		0 & 0 & 0 & \cos\theta & \sin\theta & 0 & 0 & 0 \\
		0 & 0 & 0 & -\sin\theta & \cos\theta & 0 & 0 & 0 \\
		-\sin\theta & 0 & 0 & 0 & 0 & \cos\theta & 0 & 0 \\
		0 & 0 & 0 & 0 & 0 & 0 & 1 & 0 \\
		0 & 0 & 0 & 0 & 0 & 0 & 0 & 1
	\end{array}
	\right),
\end{equation}
\endgroup
where the mixing angle satisfies
\begin{equation}
	\tan\theta=
	\frac{2\Delta_2}
	{-\Delta_1-\epsilon_{\mathrm{vb}-1}
		+\sqrt{4\Delta_2^2+(\Delta_1+\epsilon_{\mathrm{vb}-1})^2}} .
\end{equation}

The rotated basis states are
{\setlength{\jot}{2pt}
	\begin{equation}
		\begin{aligned}
			\ket{\Phi_1} &=
			\cos\theta\ket{\psi_{\mathrm{cb}},\uparrow}
			-\sin\theta\ket{\psi_{\mathrm{vb}-1},\downarrow}, \\
			\ket{\Phi_2} &=
			\ket{\psi_{\mathrm{cb}},\downarrow}, \\
			\ket{\Phi_3} &=
			\ket{\psi_{\mathrm{vb}},\uparrow}, \\
			\ket{\Phi_4} &=
			\cos\theta\ket{\psi_{\mathrm{vb}},\downarrow}
			-\sin\theta\ket{\psi_{\mathrm{vb}-1},\uparrow}, \\
			\ket{\Phi_5} &=
			\sin\theta\ket{\psi_{\mathrm{vb}},\downarrow}
			+\cos\theta\ket{\psi_{\mathrm{vb}-1},\uparrow}, \\
			\ket{\Phi_6} &=
			\sin\theta\ket{\psi_{\mathrm{cb}},\uparrow}
			+\cos\theta\ket{\psi_{\mathrm{vb}-1},\downarrow}, \\
			\ket{\Phi_7} &=
			\ket{\psi_{\mathrm{vb}-2},\uparrow}, \\
			\ket{\Phi_8} &=
			\ket{\psi_{\mathrm{vb}-2},\downarrow}.
		\end{aligned}
	\end{equation}
}

In this basis, the $\Gamma$-point Hamiltonian becomes diagonal:
\begingroup
\scriptsize
\setlength{\arraycolsep}{2.5pt}
\renewcommand{\arraystretch}{0.9}
\begin{equation}
	U^{-1}H_{8\times8}(\mathbf{0})U=
	\left(
	\begin{array}{cccccccc}
		A_1 & 0 & 0 & 0 & 0 & 0 & 0 & 0 \\
		0 & A_2 & 0 & 0 & 0 & 0 & 0 & 0 \\
		0 & 0 & A_2 & 0 & 0 & 0 & 0 & 0 \\
		0 & 0 & 0 & A_1 & 0 & 0 & 0 & 0 \\
		0 & 0 & 0 & 0 & A_3 & 0 & 0 & 0 \\
		0 & 0 & 0 & 0 & 0 & A_3 & 0 & 0 \\
		0 & 0 & 0 & 0 & 0 & 0 & A_4 & 0 \\
		0 & 0 & 0 & 0 & 0 & 0 & 0 & A_4
	\end{array}
	\right).
\end{equation}
\endgroup

The corresponding eigenvalues are
\begin{equation}
	\begin{aligned}
		A_1 &= \frac{1}{2}\left[
		\epsilon_{\mathrm{vb}-1}-\Delta_1
		+\sqrt{(\Delta_1+\epsilon_{\mathrm{vb}-1})^2+4\Delta_2^2}
		\right],\\
		A_2 &= \Delta_1,\\
		A_3 &= \frac{1}{2}\left[
		\epsilon_{\mathrm{vb}-1}-\Delta_1
		-\sqrt{(\Delta_1+\epsilon_{\mathrm{vb}-1})^2+4\Delta_2^2}
		\right],\\
		A_4 &= \epsilon_{\mathrm{vb}-2}.
	\end{aligned}
	\label{eq:A_params}
\end{equation}

Next, we transform the momentum-dependent
$\mathbf{k}\cdot\mathbf{p}$ Hamiltonian into the rotated basis. Using
$k_{\pm}=k_x\pm i k_y$, one obtains
\begin{equation}
	\resizebox{\textwidth}{!}{$
		U^{-1} H_{\mathbf{k}\cdot\mathbf{p}} U=
		\begin{pmatrix}
			0 & 0 & i\gamma_1\cos\theta\, k_- & 0 & 0 & 0 & i\gamma_2\cos\theta\, k_+ & 0 \\
			0 & 0 & 0 & i\gamma_1\cos\theta\, k_- & 0 & 0 & 0 & i\gamma_2 k_+ \\
			-i\gamma_1\cos\theta\, k_+ & 0 & 0 & 0 & 0 & 0 & -i\gamma_2 k_- & 0 \\
			0 & -i\gamma_1\cos\theta\, k_+ & 0 & 0 & 0 & 0 & 0 & -i\gamma_2\cos\theta\, k_- \\
			0 & 0 & 0 & 0 & 0 & 0 & 0 & i\gamma_2\sin\theta\, k_- \\
			0 & 0 & 0 & 0 & 0 & 0 & -i\gamma_2\sin\theta\, k_+ & 0 \\
			-i\gamma_2\cos\theta\, k_- & 0 & i\gamma_2 k_+ & 0 & 0 & i\gamma_2\sin\theta\, k_- & 0 & 0 \\
			0 & -i\gamma_2 k_- & 0 & i\gamma_2\cos\theta\, k_+ & -i\gamma_2\sin\theta\, k_+ & 0 & 0 & 0
		\end{pmatrix}.
		$}
\end{equation}

The transformed Hamiltonian is then partitioned into the low-energy subspace
spanned by $\{\ket{\Phi_1},\ket{\Phi_2},\ket{\Phi_3},\ket{\Phi_4}\}$ and the
remote subspace spanned by
$\{\ket{\Phi_5},\ket{\Phi_6},\ket{\Phi_7},\ket{\Phi_8}\}$. In block form, the
Hamiltonian can be written as
$
	H=
	\begin{pmatrix}
		\mathcal{A} & 0 \\
		0 & \mathcal{B}
	\end{pmatrix}
	+
	\begin{pmatrix}
		\mathcal{C} & \mathcal{F} \\
		\mathcal{F}^{\dagger} & \mathcal{D}
	\end{pmatrix},
$
where $\mathcal{A}$ and $\mathcal{B}$ contain the diagonal energies of the
low-energy and remote subspaces, respectively. The matrices $\mathcal{C}$ and
$\mathcal{D}$ describe the intra-subspace momentum-dependent couplings, while
$\mathcal{F}$ contains the coupling between the low-energy and remote subspaces.

Explicitly,
\begin{equation}
	\resizebox{\textwidth}{!}{$
		\begin{aligned}
			\mathcal{A} &=
			\begin{pmatrix}
				A_1 & 0 & 0 & 0 \\
				0 & A_2 & 0 & 0 \\
				0 & 0 & A_2 & 0 \\
				0 & 0 & 0 & A_1
			\end{pmatrix},
			&
			\mathcal{B} &=
			\begin{pmatrix}
				A_3 & 0 & 0 & 0 \\
				0 & A_3 & 0 & 0 \\
				0 & 0 & A_4 & 0 \\
				0 & 0 & 0 & A_4
			\end{pmatrix}, \\[2mm]
			\mathcal{C} &=
			\begin{pmatrix}
				0 & 0 & i\gamma_1\cos\theta\,k_- & 0 \\
				0 & 0 & 0 & i\gamma_1\cos\theta\,k_- \\
				-i\gamma_1\cos\theta\,k_+ & 0 & 0 & 0 \\
				0 & -i\gamma_1\cos\theta\,k_+ & 0 & 0
			\end{pmatrix},
			&
			\mathcal{D} &=
			\begin{pmatrix}
				0 & 0 & 0 & i\gamma_2\sin\theta\,k_- \\
				0 & 0 & -i\gamma_2\sin\theta\,k_+ & 0 \\
				0 & i\gamma_2\sin\theta\,k_- & 0 & 0 \\
				-i\gamma_2\sin\theta\,k_+ & 0 & 0 & 0
			\end{pmatrix}, \\[2mm]
			\mathcal{F} &=
			\begin{pmatrix}
				0 & 0 & i\gamma_2\cos\theta\,k_+ & 0 \\
				0 & 0 & 0 & i\gamma_2 k_+ \\
				0 & 0 & -i\gamma_2 k_- & 0 \\
				0 & 0 & 0 & -i\gamma_2\cos\theta\,k_-
			\end{pmatrix}.
		\end{aligned}
		$}
\end{equation}

Using second-order Löwdin partitioning, the effective Hamiltonian in the
low-energy subspace is
\begin{equation}
	H^{\mathrm{eff}}_{mm'}=
	\mathcal{A}_{mm'}+\mathcal{C}_{mm'}
	+\frac{1}{2}\sum_l
	\mathcal{F}_{ml}\mathcal{F}_{m'l}^{*}
	\left[
	\frac{1}{\mathcal{A}_{mm}-\mathcal{B}_{ll}}
	+
	\frac{1}{\mathcal{A}_{m'm'}-\mathcal{B}_{ll}}
	\right].
\end{equation}

Keeping terms up to second order in $\mathbf{k}$, the effective 4 $\times$ 4
Hamiltonian becomes
\begingroup
\setlength{\arraycolsep}{3pt}
\renewcommand{\arraystretch}{0.95}
\begin{equation}
	H_{4\times4}(\mathbf{k})=
	\begin{pmatrix}
		\mathcal{H}_1(\mathbf{k}) & 0 & \mathcal{H}_3(\mathbf{k}) & 0 \\
		0 & \mathcal{H}_2(\mathbf{k}) & 0 & \mathcal{H}_3(\mathbf{k}) \\
		\mathcal{H}_3^{*}(\mathbf{k}) & 0 & \mathcal{H}_2(\mathbf{k}) & 0 \\
		0 & \mathcal{H}_3^{*}(\mathbf{k}) & 0 & \mathcal{H}_1(\mathbf{k})
	\end{pmatrix},
\end{equation}
\endgroup

with
$
	\mathcal{H}_{1,2}(\mathbf{k})=A_{1,2}+B_{1,2}k^2,
	\
	\mathcal{H}_3(\mathbf{k})=iNk_-+B_3k_+^2 .
$
The coefficients are
{\setlength{\jot}{2pt}
	\begin{equation}
		\begin{aligned}
			N &= \gamma_1\cos\theta, \\
			B_1 &= \frac{\hbar^2}{2m_0} + \frac{\gamma_2^2\cos^2\theta}{A_1-A_4}, \\
			B_2 &= \frac{\hbar^2}{2m_0} + \frac{\gamma_2^2}{A_2-A_4}, \\
			B_3 &= -\frac{\cos\theta}{2}
			\left(\frac{\gamma_2^2}{A_1-A_4} + \frac{\gamma_2^2}{A_2-A_4}\right).
		\end{aligned}
	\end{equation}
}

The quadratic off-diagonal term $B_3k_+^2$ originates from virtual transitions
through the eliminated $\mathrm{VB}-2$ states. It is therefore a direct
consequence of the Löwdin downfolding procedure and encodes the leading
second-order coupling between the low-energy conduction and valence sectors.

Finally, by reordering the low-energy basis as
$\{\ket{\Phi_1},\ket{\Phi_3},\ket{\Phi_2},\ket{\Phi_4}\}$, the Hamiltonian can
be written as two time-reversed $2\times2$ blocks. One block is
\begin{equation}
	H_{+}(\mathbf{k})=
	\begin{pmatrix}
		\mathcal{H}_1(\mathbf{k}) & \mathcal{H}_3(\mathbf{k}) \\
		\mathcal{H}_3^{*}(\mathbf{k}) & \mathcal{H}_2(\mathbf{k})
	\end{pmatrix},
\end{equation}
while the other block, $H_{-}(\mathbf{k})$, is related to $H_{+}(\mathbf{k})$ by
time-reversal symmetry and therefore has the same eigenvalues. For the fitted
parameters of OsC and RuC, the linear coupling parameter $N$ is numerically very
small. In the parameter regime relevant to OsC and RuC, the fitted value of the linear
coupling $N$ is much smaller than the quadratic downfolding contribution.
Neglecting this small linear term gives the dispersion relation used in the main
text, Eq.~\eqref{eq:four_band_eigenvalues}.

\bibliography{baradaran_ref_cleaned_partial}

\end{document}